\documentclass[11pt,preprint]{aastex}

\usepackage{graphicx,amsmath,booktabs,threeparttable,url}

\usepackage[breaklinks]{hyperref}
\usepackage[hyphenbreaks]{breakurl}

\usepackage{CJKutf8}
\usepackage{threeparttable}
\usepackage[usenames,dvips]{color}
\usepackage[normalem]{ulem}
\usepackage{subfigure}

\newcommand{\cntext}[1]{\begin{CJK*}{UTF8}{bsmi}#1\end{CJK*}}

\newcommand{\Teff}{\ensuremath{T_{\mathrm{eff}}}}

\usepackage{natbib}

\begin{document}

\title{SEARCH FOR SURVIVING COMPANIONS IN TYPE Ia SUPERNOVA REMNANTS }

\author{Kuo-Chuan Pan (\cntext{潘國全})$^{1}$, Paul M. Ricker$^2$, and Ronald E. Taam$^{3,4}$}
\affil{$^1$Physik Department, Universit\"{a}t Basel, Klingelbergstrasse 82, CH-4056 Basel, Switzerland; kuo-chuan.pan@unibas.ch}
\affil{$^2$Department of Astronomy, University of Illinois at Urbana$-$Champaign, 1002 West Green Street, Urbana, IL 61801, USA; pmricker@illinois.edu}
\affil{$^3$Department of Physics and Astronomy, Northwestern University, 2145 Sheridan Road, Evanston, IL 60208, USA; r-taam@northwestern.edu}
\affil{$^4$Academia Sinica Institute of Astronomy and Astrophysics, P.O. Box 23-141, Taipei 10617, Taiwan; taam@asiaa.sinica.edu.tw}


\begin{abstract}
The nature of the progenitor systems of type~Ia supernovae is still unclear.
One way to distinguish between the single-degenerate scenario and
double-degenerate scenario for their progenitors is to search for the surviving companions.
Using a technique that couples the results from multi-dimensional hydrodynamics
simulations with calculations of the structure and evolution of main-sequence-
and helium-rich surviving companions, the color and magnitude of 
main-sequence- and helium-rich surviving companions are predicted as functions of time.  
The surviving companion candidates in Galactic type~Ia supernova remnants
and nearby extragalactic type~Ia supernova remnants are discussed.
We find that the maximum detectable distance of 
main-sequence surviving companions (helium-rich surviving companions) 
is $0.6-4$~Mpc ($0.4-16$~Mpc),
if the apparent magnitude limit is 27 in the absence of extinction,
suggesting that the Large and Small Magellanic Clouds and the Andromeda Galaxy are
excellent environments in which to search for surviving companions.
However, only five Ia~SNRs have been searched for surviving companions, 
showing little support for the standard channels in the singe-degenerate scenario. 
To better understand the progenitors of type Ia supernovae, 
we encourage the search for surviving companions in other nearby 
type Ia supernova remnants.
\end{abstract}



\section{Introduction}

Thermonuclear explosions of carbon-oxygen (CO) white dwarfs (WDs),
which lead to Type~Ia supernovae (SNe~Ia),
could be triggered by the merger of two WDs
(the double-degenerate scenario (DDS) \citealt{1984ApJS...54..335I, 1984ApJ...277..355W})
or by the accretion of matter from a non-degenerate star through Roche-lobe overflow
(the single-degenerate scenario (SDS) \citealt{1973ApJ...186.1007W, 1982ApJ...257..780N}).
In the SDS, the companion to the CO-WD could be a main-sequence (MS),
red giant (RG), or helium-rich (He) star. In the DDS,
the companion could be another CO~WD or He~WD.
Based on current studies, both scenarios are not ruled out by observations, 
but are also not proven by them. 
If both scenarios lead to SN~Ia,
it is still unclear which channel(s) in the SDS and DDS dominate(s) the SNe~Ia,
and by what fraction \citep{2012NewAR..56..122W, 2013FrPhy...8..116H, 2013arXiv1312.0628M, 2014arXiv1403.4087R}.

Recent multi-dimensional hydrodynamics simulations of
SN impact on the non-degenerate binary companions in the SDS
\citep{2000ApJS..128..615M, 2008A&A...489..943P, 2010ApJ...715...78P, 2012ApJ...750..151P, 2012A&A...548A...2L, 2013ApJ...774...37L}
suggest that the companions should survive the SN impact and could be detectable.
Therefore, one of the direct methods to distinguish between
the SDS and DDS is to search for the surviving companions (SCs) in SN~Ia remnants (Ia~SNRs).

\citet[][hereafter P10]{2010ApJ...715...78P} and \citet[][hereafter P12a]{2012ApJ...750..151P}
examine the effects of a SN impact on the non-degenerate
binary companions in the SDS for MS, RG, and He star binary companions via multi-dimensional hydrodynamics simulations.
These simulations include the symmetry-breaking effects of orbital motion, rotation of the binary companion,
and Roche-lobe overflow (RLOF), all of which allow a better description of SN-driven shock compression, 
heating, and stripping of SCs.
\cite{2012A&A...548A...2L, 2013ApJ...774...37L} examined a similar SN impact on MS and He stars
with the SPH approach using companion models from more sophisticated one-dimensional binary evolutions.
However Liu et al. did not study the subsequent post-impact evolution
and therefore could not predict the properties of SCs in historical Ia~SNRs.
On the other hand, \cite{2003astro.ph..3660P} and \cite{2013ApJ...765..150S} examined the evolution of
a $1M_\odot$ subgiant and a $1M_\odot$ MS companion with ad hoc prescriptions
for energy input and mass stripping without performing detailed hydrodynamical calculations, thereby dramatically overestimating the luminosity of SCs.
In \citet[][hereafter P12b]{2012ApJ...760...21P} and \citet[][hereafter P13]{2013ApJ...773...49P},
we mapped our detailed three-dimensional
hydrodynamical results into a one-dimensional stellar evolution code to simulate the post-impact evolution of MS- and He-SCs and thus
provided a more realistic treatment of post-impact evolution.

In this paper, we calculate the time evolution of the magnitudes and colors of our models and discuss
the possibility of searching for SDS-SCs in specific nearby Ia~SNRs.  We present numerical results in \S~2 and compare them with searches for SCs in Galactic Ia SNRs in \S~3, in the Magellanic
Clouds in \S~4, and in M31 and M82 in \S~5.  Finally, we discuss the evidence for the SDS channel for SNe~Ia and
present our conclusions in the last section.


\section{Numerical Results and Predictions \label{sec_predictions}}

In this section, we discuss and predict some possible observables of SCs in nearby Ia~SNRs 
and describe differences between SC candidates and unrelated stars in Ia~SNRs. 
In particular, we calculate the colors, magnitudes, effective temperatures, and photospheric radii of
MS- and He-SCs as functions of time. 
We also predict the magnitude and effective temperature changes as functions of time.
The linear and rotational velocities of SCs are also discussed.
We also suggest upper limits for the Ni/Fe contamination in SCs.

\subsection{Colors and Magnitudes of Surviving Companions}

In simulations of MS- and He-SCs, the bolometric luminosity, effective temperature,
and photospheric radius are directly determined using the stellar evolution
code MESA\footnote{\url{http://mesa.sourceforge.net}} (Modules for Experiments in Stellar Astrophysics;
\citealt{2011ApJS..192....3P, 2013ApJS..208....4P}).
To facilitate direct comparison with optical observations, the bolometric luminosity is converted
to broad band magnitudes. Additional observable quantities such as the strength of
absorption lines require the treatment of detailed radiation transfer effects in stellar atmosphere models,
which are not considered in this paper.

Given the effective temperature and photospheric radius of a SC, the magnitude of the SC
can be estimated under the assumption that the photosphere emits a blackbody radiation spectrum.
We have considered several different filters with their corresponding sensitivity functions in this study,
including the Johnson-Cousins-Glass UBVIR system and the HST/WFC3 system.
The absolute magnitudes are calculated in the AB magnitude system.
For a given extinction, $A_V$, the extinction curve can be calculated using the fitting formula in
\cite{1994ApJ...422..158O}.

\begin{figure}
\plotone{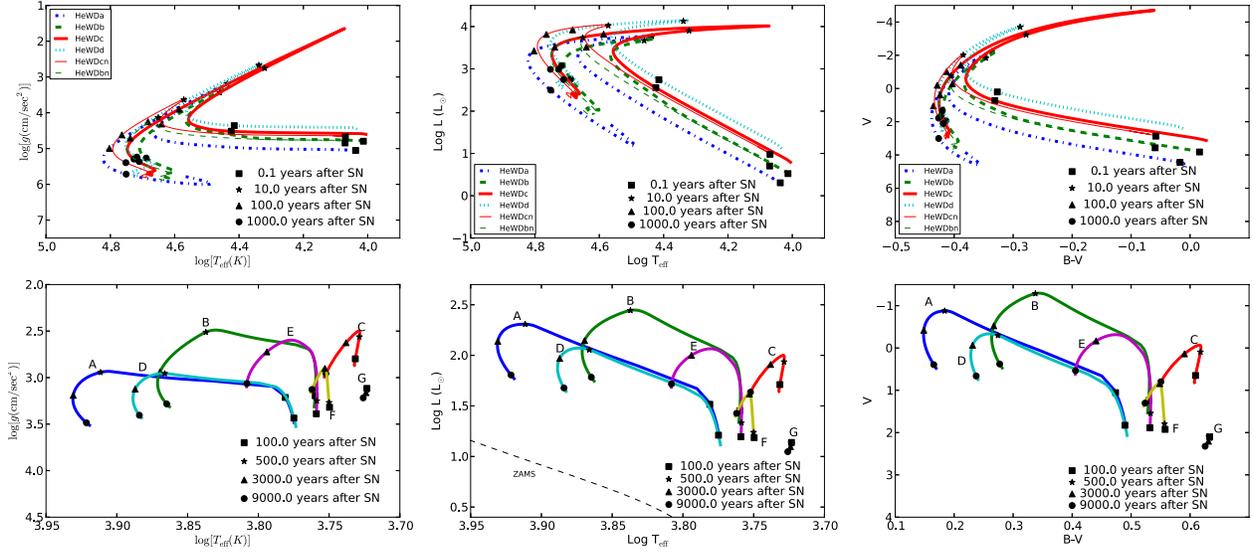}
\caption{\label{fig_hr_all}
Evolutionary tracks in H-R diagrams for
He- (upper panels) and MS-SCs (lower panels) using different representations.
Color lines indicate different progenitor companions in Table~\ref{tab_models}.
Different symbols show the evolutionary stage at different times.
}
\end{figure}

\begin{figure}
\begin{center}
\epsscale{0.45}
\plotone{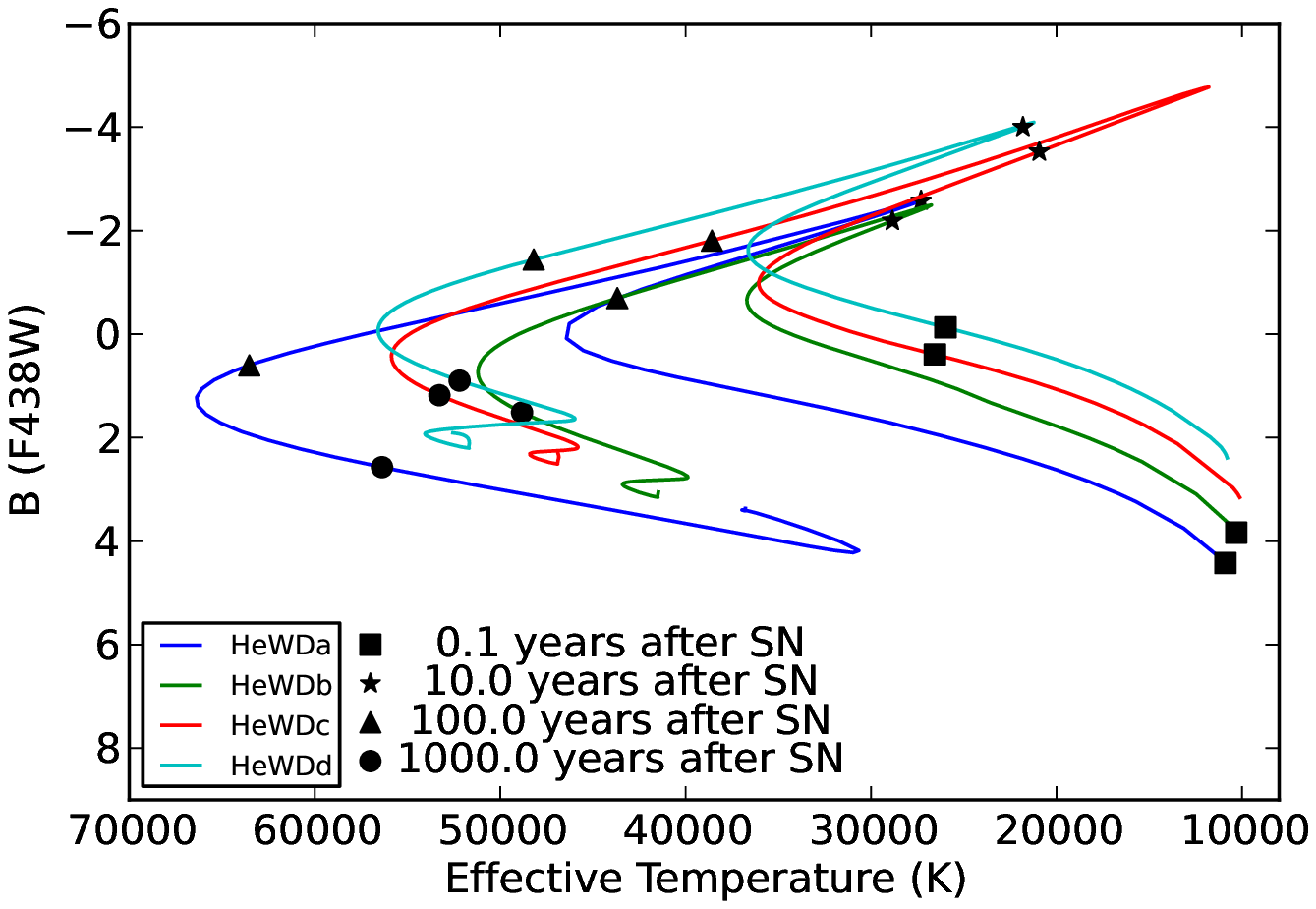}
\plotone{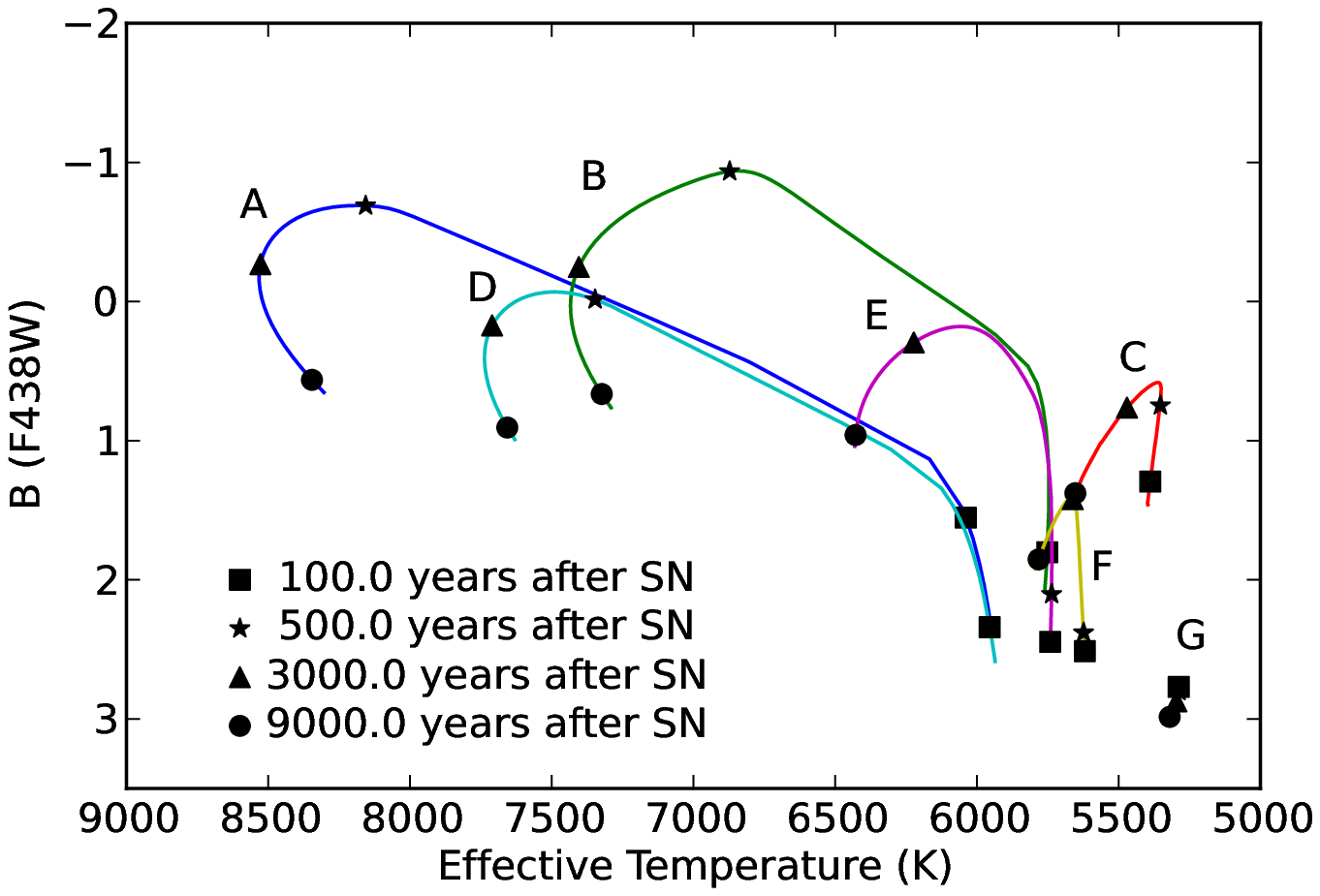}
\plotone{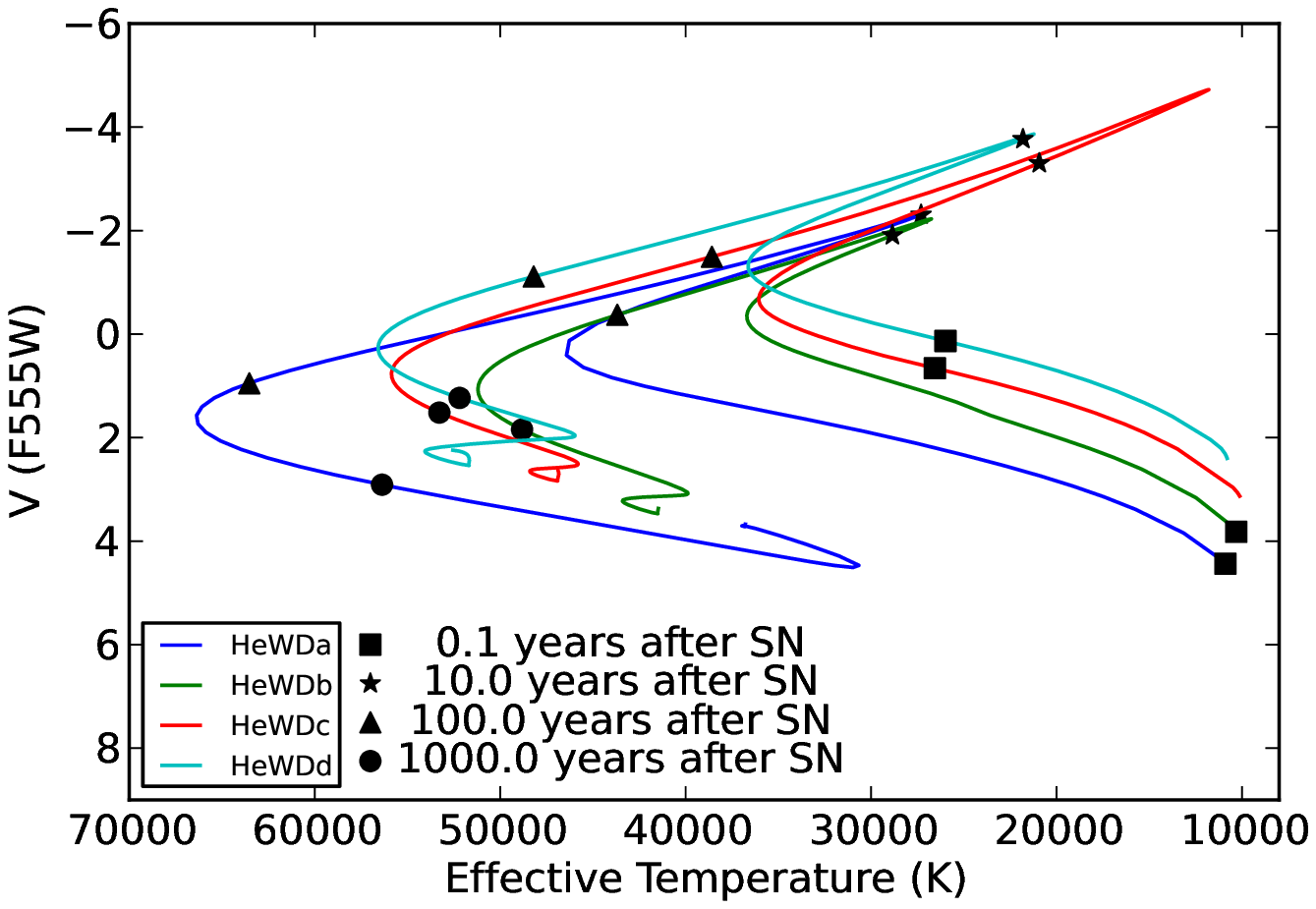}
\plotone{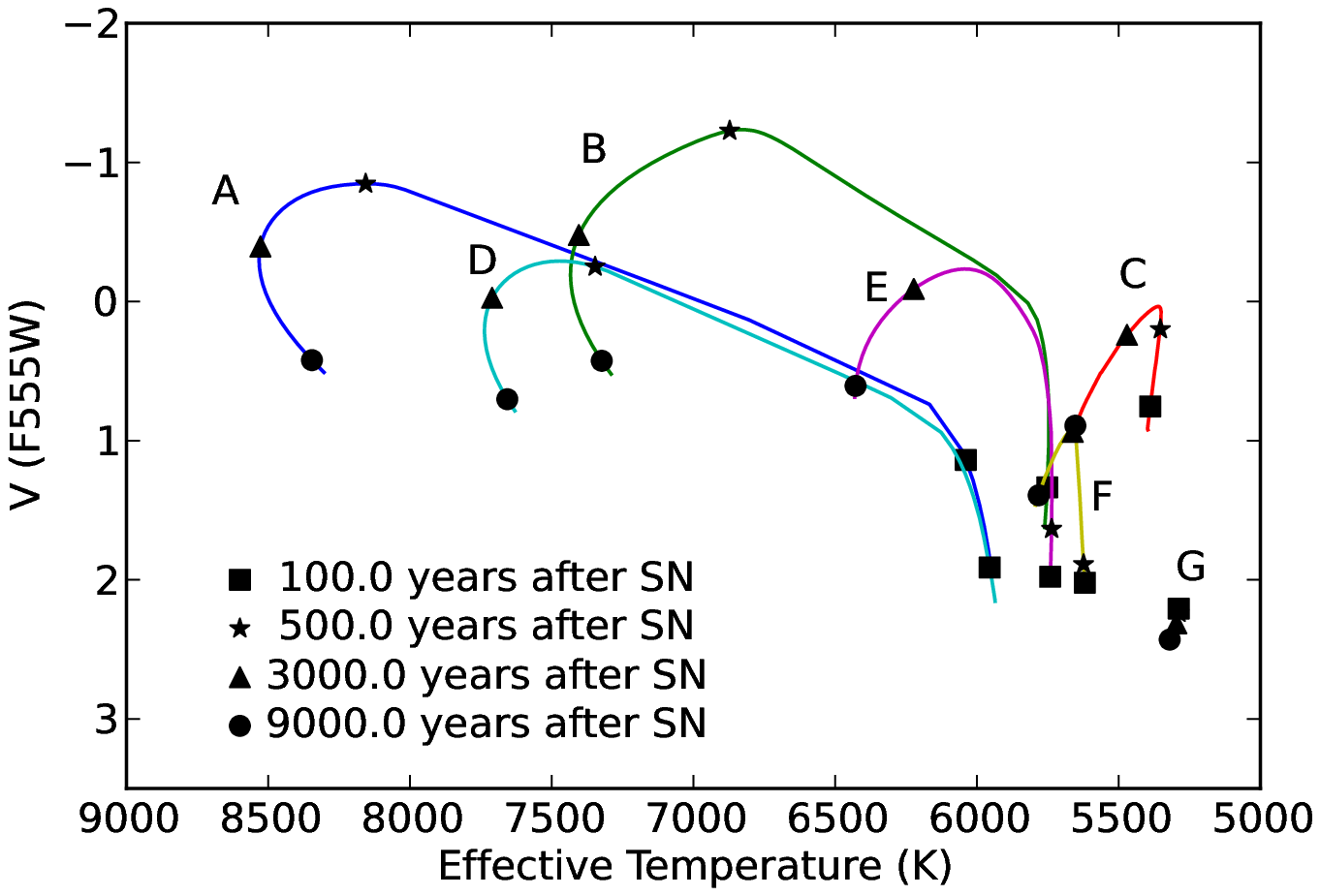}
\plotone{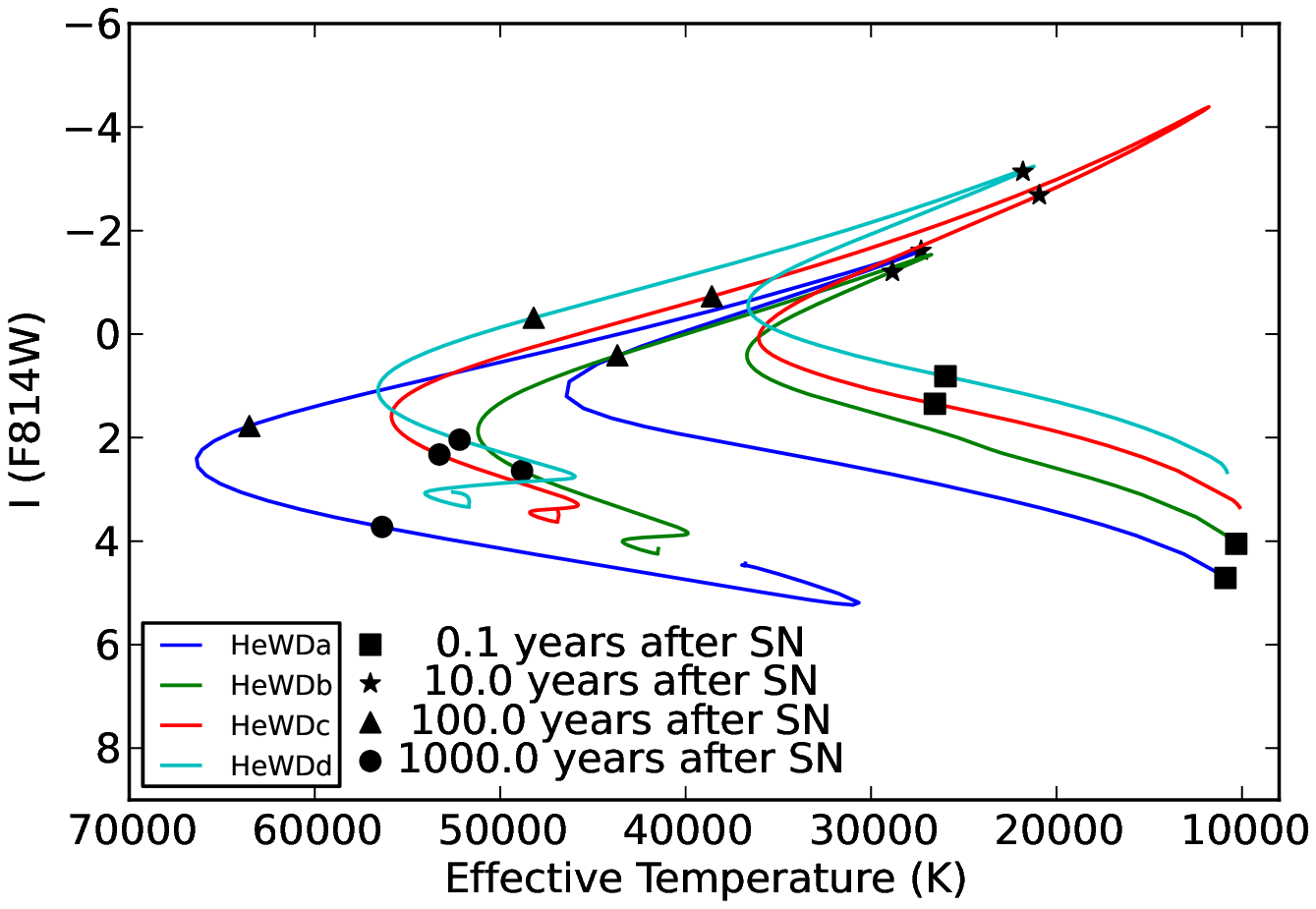}
\plotone{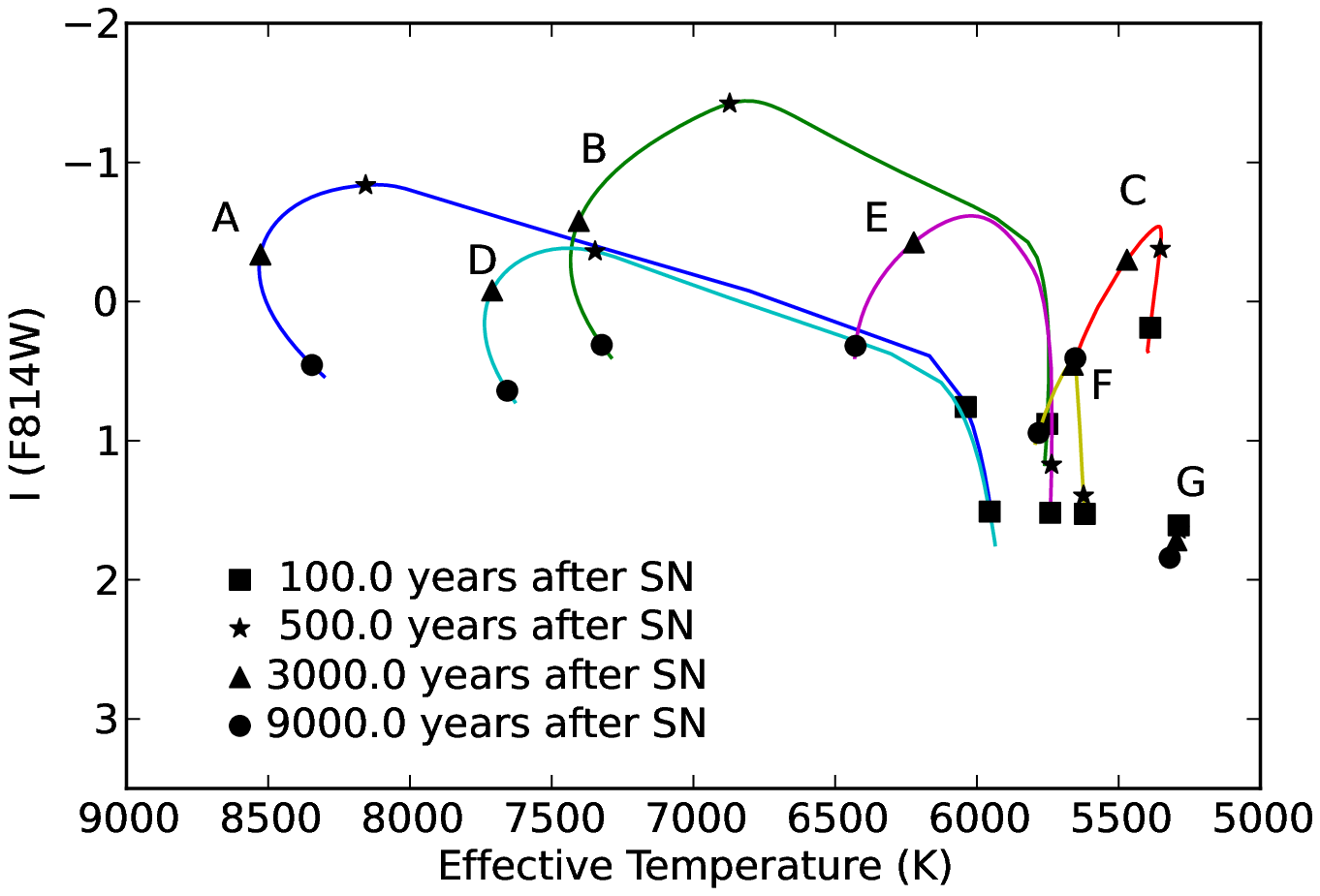}

\end{center}
\caption{\label{fig_hr_bvi}
Similar to Figure~\ref{fig_hr_all} but for different HST/WFC3 band magnitude vs.\ effective temperature.}
\end{figure}

The stellar parameters of MS- and He-SC models in our previous work are summarized in
Table~\ref{tab_models}.  Figures~\ref{fig_hr_all} and \ref{fig_hr_bvi} illustrate the
Hertzsprung-Russell (H-R) diagram in different representations using the B and V wavebands and
HST wavebands respectively.  He-SCs are sdO-like stars that exhibit stronger emission in the
$U$ and $B$ bands, while MS-SCs exhibit greater emission in the $V$ and $I$ bands (A-K subgiants).
The absolute magnitudes of MS-SCs (He-SCs) span the range
$3 < M_V \lesssim -1$~$(4 \lesssim M_B \lesssim -4)$.
The brightest phase of MS-SCs corresponds to Ia~SNR ages of $\sim 500-3000$~yr,
which are similar to the ages of most known historic nearby Ia~SNRs (see Table~\ref{tab1}). 
Therefore, if these SNRs originated from normal SNe~Ia via the SDS MS or the He star channels,
SCs should be detectable.
In the RG donor channel, almost the entire envelope of a RG should be removed during the SN impact
(\citealt{2000ApJS..128..615M}; P12a).
Therefore, the SC in Ia~SNR will no longer be a giant star,
but could be a helium degenerate core star with a shallow hydrogen-rich envelope.

It should be noted that in our calculations, we assume normal SNe~Ia
with Chandrasekhar mass explosions, and the explosions are initially spherically symmetric.
Asymmetric explosions, sub-Chandrasekhar, or super-Chandrasekhar
mass explosions may reduce or enhance the evolution of post-impact luminosity (P12b).

\subsection{Rates of Magnitude and Effective Temperature Change}

\begin{figure}
\begin{center}
\epsscale{0.8}
\plotone{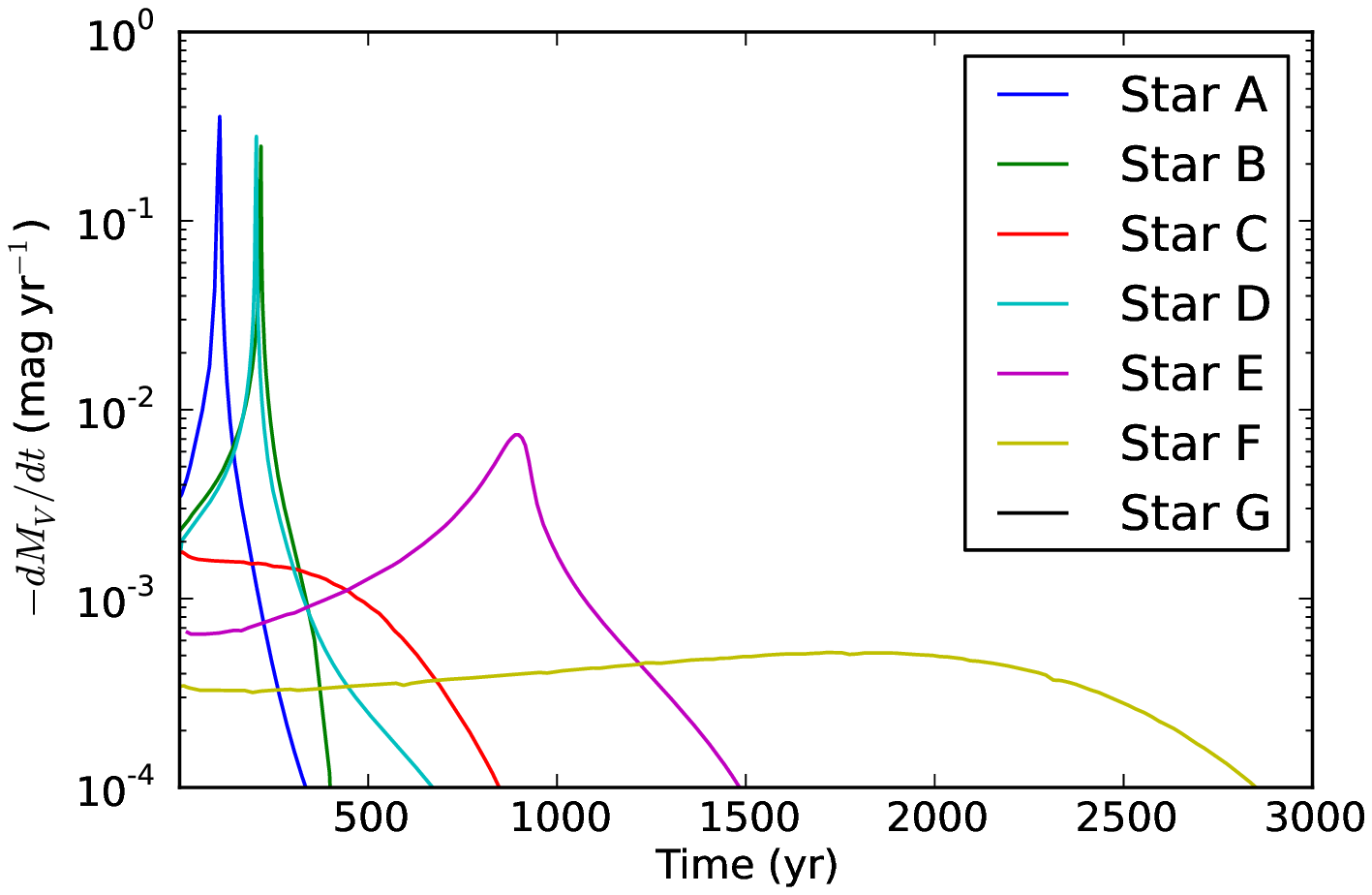}
\plotone{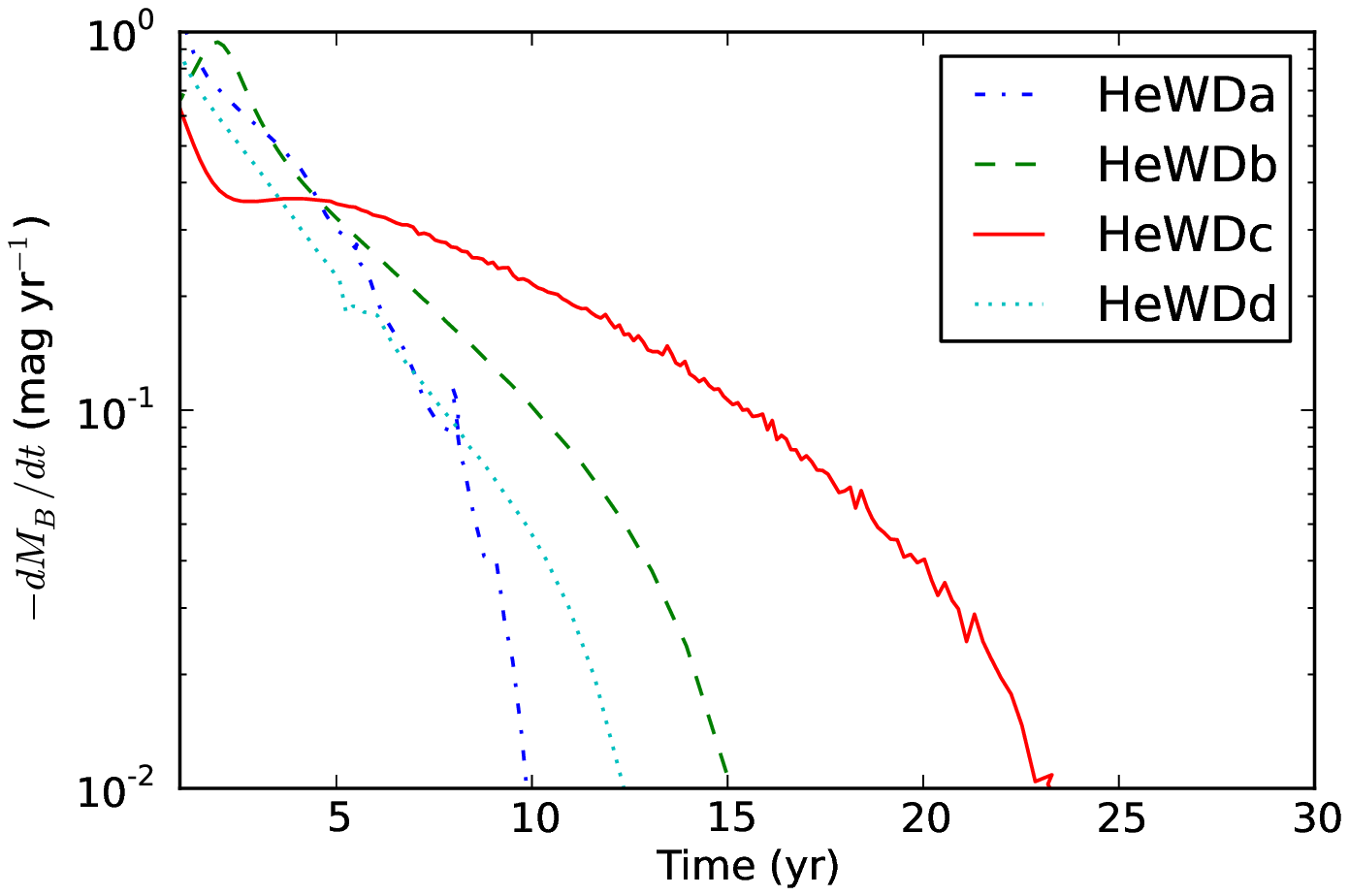}
\end{center}
\caption{\label{fig_mag}
Time derivative of magnitude change as functions of time for MS- (top) and He-SCs (bottom).
The magnitude of Star~G is increased in time and has a magnitude gradient ($dM_v/dt < 10^{-4}$), 
and therefore not shown in the figure.}
\end{figure}

\begin{figure}
\begin{center}
\epsscale{0.8}
\plotone{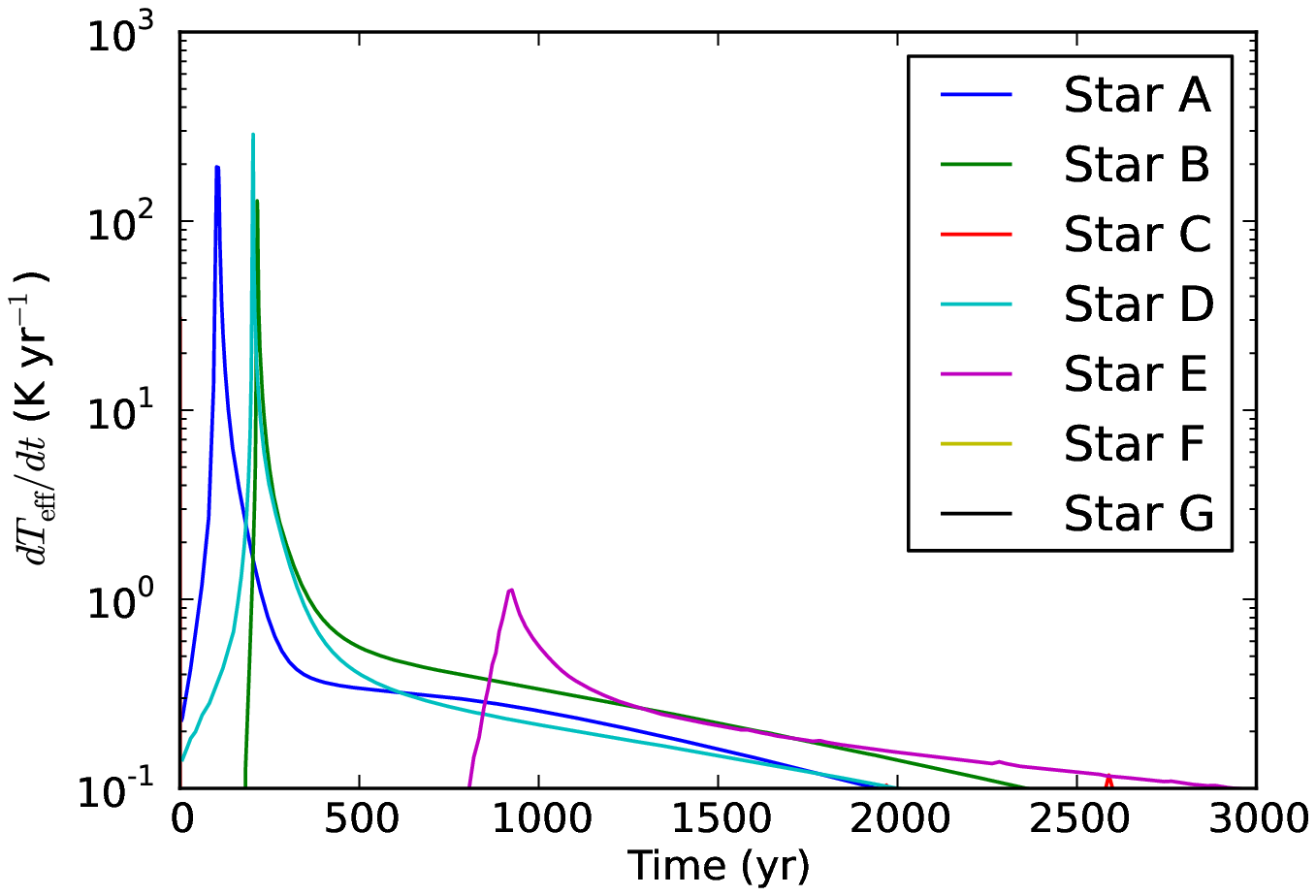}
\plotone{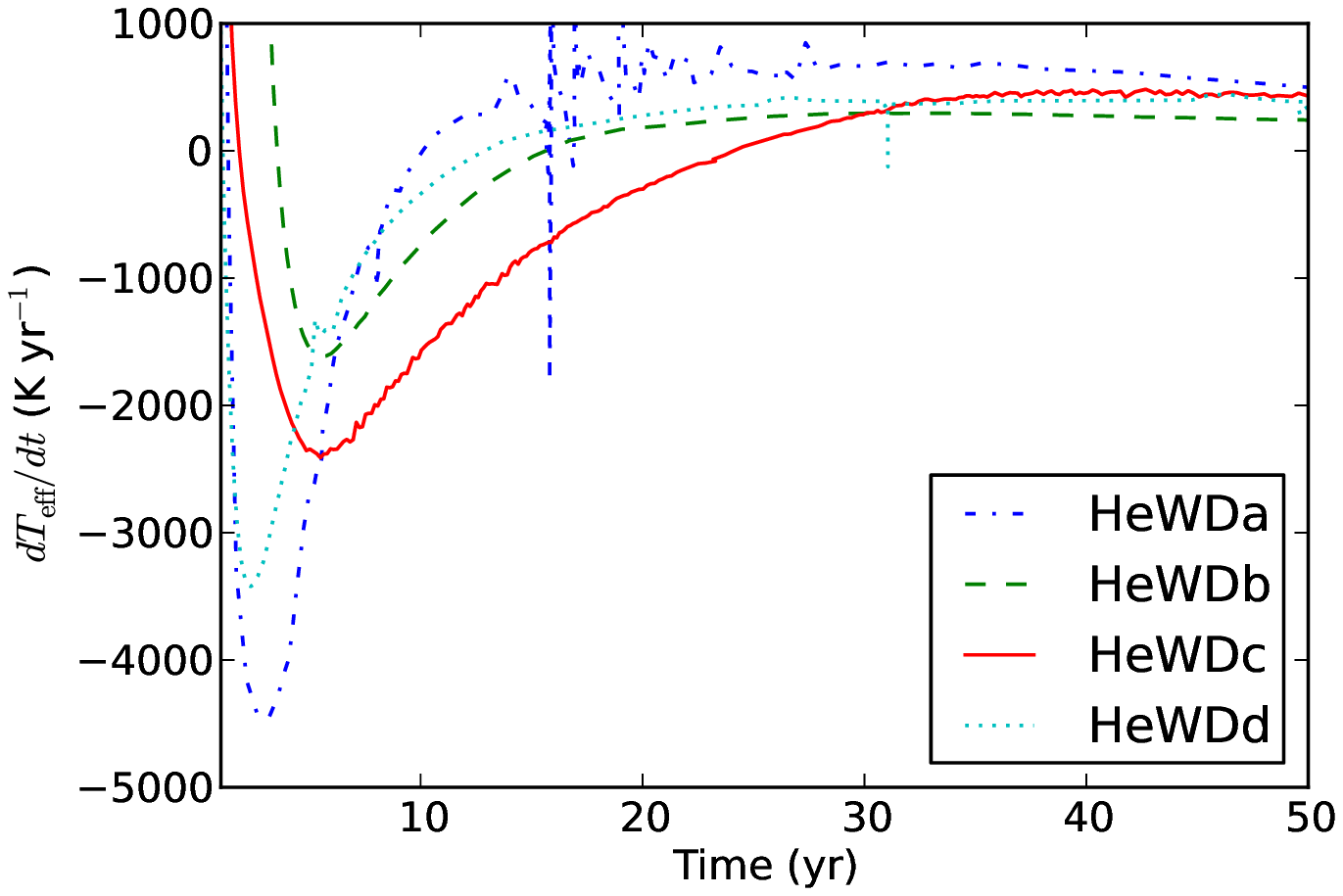}
\end{center}
\caption{\label{fig_teff}
Similar to Figure~\ref{fig_mag} but for time derivative of effective temperature.
Model~C, F, and G have change rate less than $10^{-1}$~K~yr$^{-1}$.}
\end{figure}

In addition to comparisons of the colors and magnitudes of SC candidates, 
the long term variation of magnitude and effective temperature changes is an alternative way 
to identify the SC. In Figure~\ref{fig_hr_all} we see that MS-SCs show rapid luminosity changes
but maintain similar effective temperatures in the early few hundred years after an explosion.
Therefore, a slight magnitude drop but without color change for a few years after the first observation
is expected for MS-SCs. 
On the other hand, He-SCs show both magnitude and color changes in the first ten years.  
Figure~\ref{fig_mag} and Figure~\ref{fig_teff} show the magnitude and effective temperature variations 
of our MS-  and He-SCs as functions of time.
Brighter SC models (Models~A, B, and D) have a maximum magnitude change rate
of 0.3~mag~yr$^{-1}$ in the V-band when their Ia~SNRs have ages less than five hundred years. 
For Models~C, E, and F, the MS-SCs show a smoother rate of change of magnitude during the first thousand years,
but the maximum rate is less than 0.01~mag~yr$^{-1}$, 
which is very difficult to detect with current optical telescopes. 
Model~G radiates the deposited energy immediately and shows a positive magnitude gradient (gets dimmer).
The rate of change of magnitude for Model~G is less than $10^{-4}$~mag~yr$^{-1}$.    
The effective temperatures of all our MS-SC models are roughly constants. 
The highest rate of change is about $200$~K~yr$^{-1}$ for Models A, B, and D 
and $< 1$~K~yr$^{-1}$ for  other models. 

However, the rates of change of magnitude and effective temperature for He-SCs could be notable. 
All our He-SCs change by $0.5-1$~mag~yr$^{-1}$ in the B-band during the first three years
and maintain $\sim 0.1 -0.3$~mag~yr$^{-1}$ over the first decade.  
Furthermore, He-SCs increase in temperature by $\sim 1,000$~K~yr$^{-1}$ during the first $1-3$ years
and then decrease by  $-1,000 \sim -4,500$~K~yr$^{-1}$ in the first decade. 
Therefore, these magnitude and effective temperature changes could be detected in young and nearby Ia~SNRs
if the SC is a He star.

\subsection{Stellar Velocity and Search Radius}

A SC will have an abnormal speed due to its original orbital speed plus the SN kick.
The momentum transfer between the SN ejecta with the donor star will give a kick velocity
perpendicular to its orbital velocity.
The kick velocity is about $\sim 50-100$~km~s$^{-1}$ for MS and He star donor channels,
and $\lesssim 30$ ~km~s$^{-1}$ for the RG donor channel
(P10; P12a; \citealt{2012A&A...548A...2L, 2013ApJ...774...37L}).
By adding the kick velocity to the orbital velocity,
the final linear speed at the end of supernova impact is $\sim 130- 270$~km~s$^{-1}$
for MS-SCs, $\sim 440- 730$~km~s$^{-1}$ for He-SCs (see Table~\ref{tab_models}),
and $\lesssim 50$~km~s$^{-1}$ for RG-SCs (P12a).
The contribution from the kick velocity could be as high as $30\%$ in the MS donor channel.
Assuming the CO~WD has a similar orbital speed (but opposite direction)
at the time of explosion and the SNR is moving with this orbital speed,
the observational error circle has to include a radius of at least $R_{\rm obs} > 2 v_{\rm linear} t_{\rm SNR} \sin i$,
where $t_{\rm SNR}$ is the age of the SNR, and $i$ is the inclination angle.
The theoretical estimate of the maximum $v_{\rm linear}$ is $\sim 1,000$~km~s$^{-1}$
for the He star donor channel \citep{2009A&A...508L..27W}.

\subsection{Surface Rotational Speed}

The surface rotational speed of a SC also could be an important signature
for SC searches in nearby Ia~SNRs.
At the time of the SN~Ia explosion, the companion should be synchronized by tidal locking
and be rapidly rotating, resulting in a surface rotational speed close to $\sim 100$~km~s$^{-1}$
for MS donors, $\sim 200$~km~s$^{-1}$ for He star donors, and $\lesssim 50$~km~s$^{-1}$ for RG donors
(P12a; \citealt{2011ScChG..54.2296M, 2013A&A...554A.109L}).
However, because of the angular momentum loss accompanying mass stripping,
the surface rotational speed could drop to $\sim 5-30$~km~s$^{-1}$ for MS donors
and $\sim 10-50$~km~s$^{-1}$ for He star donors after the SN impact (P12b; P13).
Once the SN ejecta escape and the SC reaches hydrostatic equilibrium,
the rotational speed could keep decreasing or increasing as a result of post-impact contraction or expansion.
If we assume the specific angular momentum is conserved during the post-impact evolution,
the surface rotational speed could be in the range from $\lesssim 10$~km~s$^{-1}$ to $\sim 200$~km~s$^{-1}$,
depending on the amount and depth of SN energy deposition
and the age of the SNR (see Figure~11 in P12b and Figure~12 in P13).

In P12b and P13, we have shown that the surface rotation speed will first drop to $\sim 10-30$~km~s$^{-1}$
after the SN impact and then keep decreasing to $< 10$~km~s$^{-1}$ during the early expansion
over a timescale of $10^{2.5}-10^3$~yr for MS-SCs and $10-30$~yr for He-SCs.
The stellar expansion is due to the release of energy deposited by the SN ejecta.
Once the deposited energy has all been released, the envelope of the SC will contract, releasing its gravitational energy.
In this phase, the rotational speed may increase up to $10-30$~km~s$^{-1}$ for MS-SCs
and $\sim 200$~km~s$^{-1}$ for He-SCs.
For most of the lifespan of Ia~SNRs with ages less than a thousand years,
MS-SCs are slowly rotating subgiants but He-SCs are rapidly rotating sdO/B stars.

\subsection{Ejecta Contamination}

\cite{2006ApJ...644..954O} have suggested that the contamination from SN ejecta may provide
observable features in iron absorption lines, which can be used to identify the SCs in Ia~SNRs.
In P12a, we showed that the amount of Ni/Fe contamination at the surfaces of the SCs
is $\sim 10^{-5} M_\odot$ for MS star companions, $\sim 10^{-4} M_\odot$ for He star companions,
and $\sim 10^{-8} M_\odot$ for RG companions. At the current stage, our simplified post-impact evolution method
cannot predict the strength of any absorption or emission lines during the evolution, but we can provide
an order-of-magnitude estimate by assuming that these contaminated material are uniformly mixed in the stellar envelope.
The estimated upper limit for the nickel-to-hydrogen-plus-helium ratio is about $\sim 10^{-4}$ for MS-SCs,
$\sim 10^{-3}$ for He-SCs, and $\sim 10^{-5}$ for RG-SCs (P12a).

\section{Galactic Type Ia SNRs \label{sec_milkyway}}

The rate of Galactic SNe (including Type I and Type II) is about $2.5^{+0.8}_{-0.5}$~SNe per century,
and $\sim 15\%$ of them are SNe~Ia \citep{1994ApJS...92..487T}.
These rates suggest that there should be more than 
2,500 SNRs in our Galaxy, and 300 of them are Ia~SNRs,
if SNRs are recognizable for ages less than $10^5$~yrs.  
However, only $\sim 312$ Galactic SNRs have been identified
\citep{2012AdSpR..49.1313F}, and only four of them are known as (or likely to be) Ia~SNRs ($< 2\%$).
Therefore, a search for a SC in the central region of a SNR could also be an alternative indirect method
to identify Ia~SNRs among currently-known SNRs.
Table~\ref{tab1} shows a summary of the four possible Galactic Ia~SNRs (and/or Ia~SNR candidates):
SN~1006, SN~1572, SN~1604, and RCW~86.

\subsection{SN~1572 (Tycho's SNR)}

The young ($442$~yr) and nearby ($2.8 \pm 0.4$~kpc, \citealt{2004ApJ...612..357R}) SNR,
Tycho's SNR, has been identified as a normal Ia~SNR by its scattered-light echo \citep{2008Natur.456..617K}.
The non-thermal X-ray arc within Tycho's SNR could arise from an interaction between the SN ejecta
and mass ejected from the companion star, giving support for the SDS as the progenitor \citep{2011ApJ...732...11L}.

\begin{figure}
\begin{center}
\epsscale{0.5}
\plotone{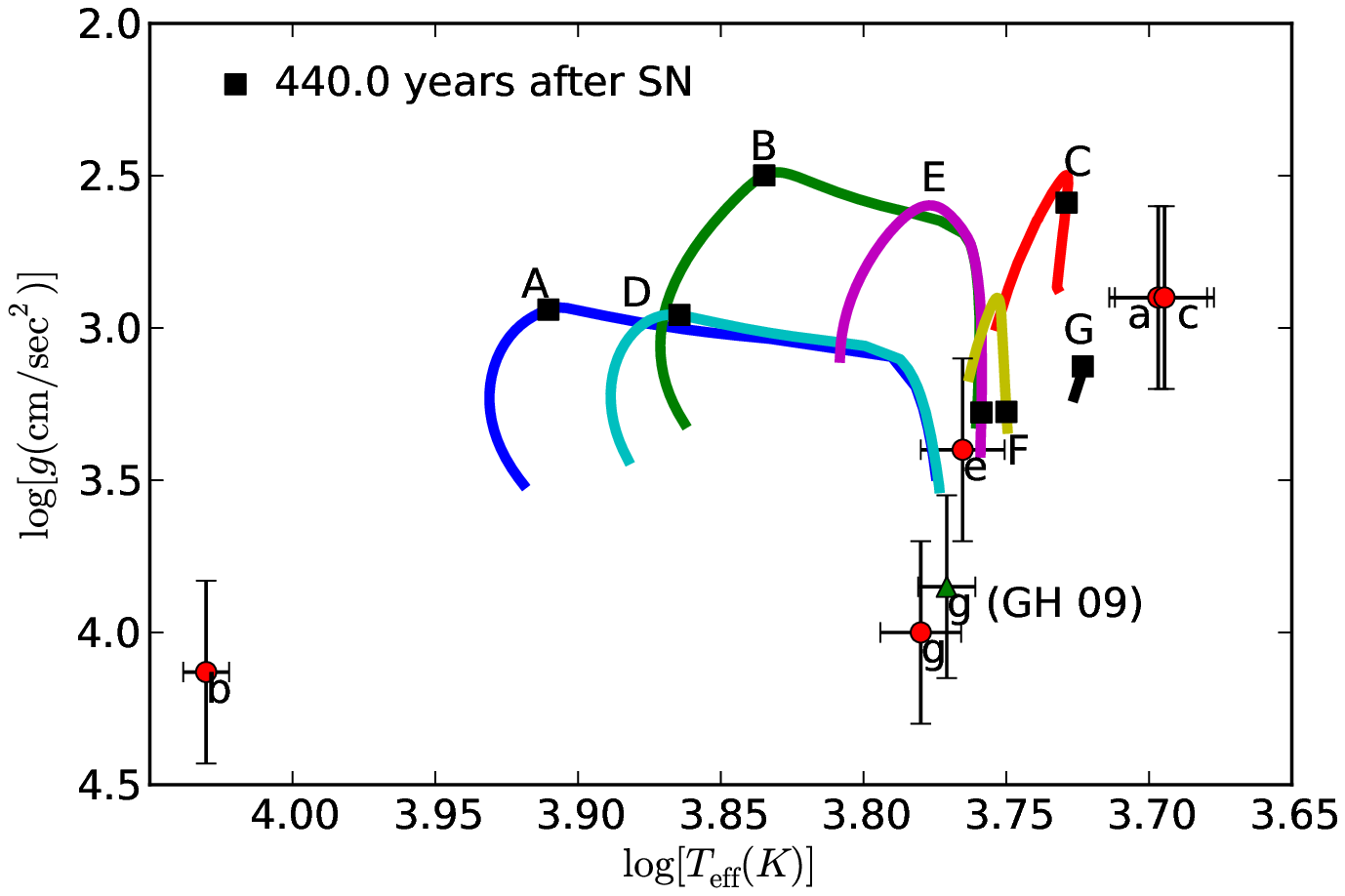}
\plotone{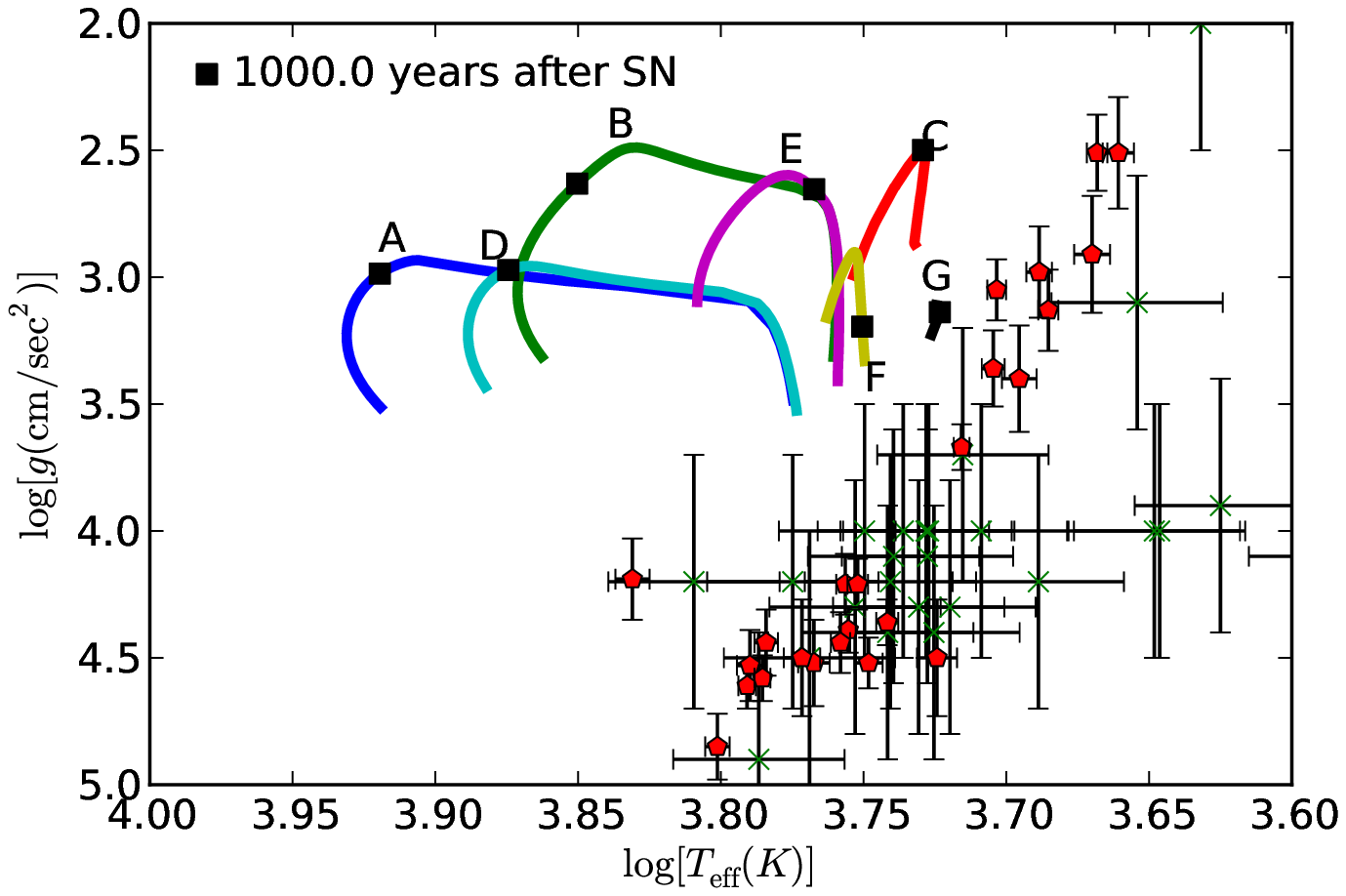}
\plotone{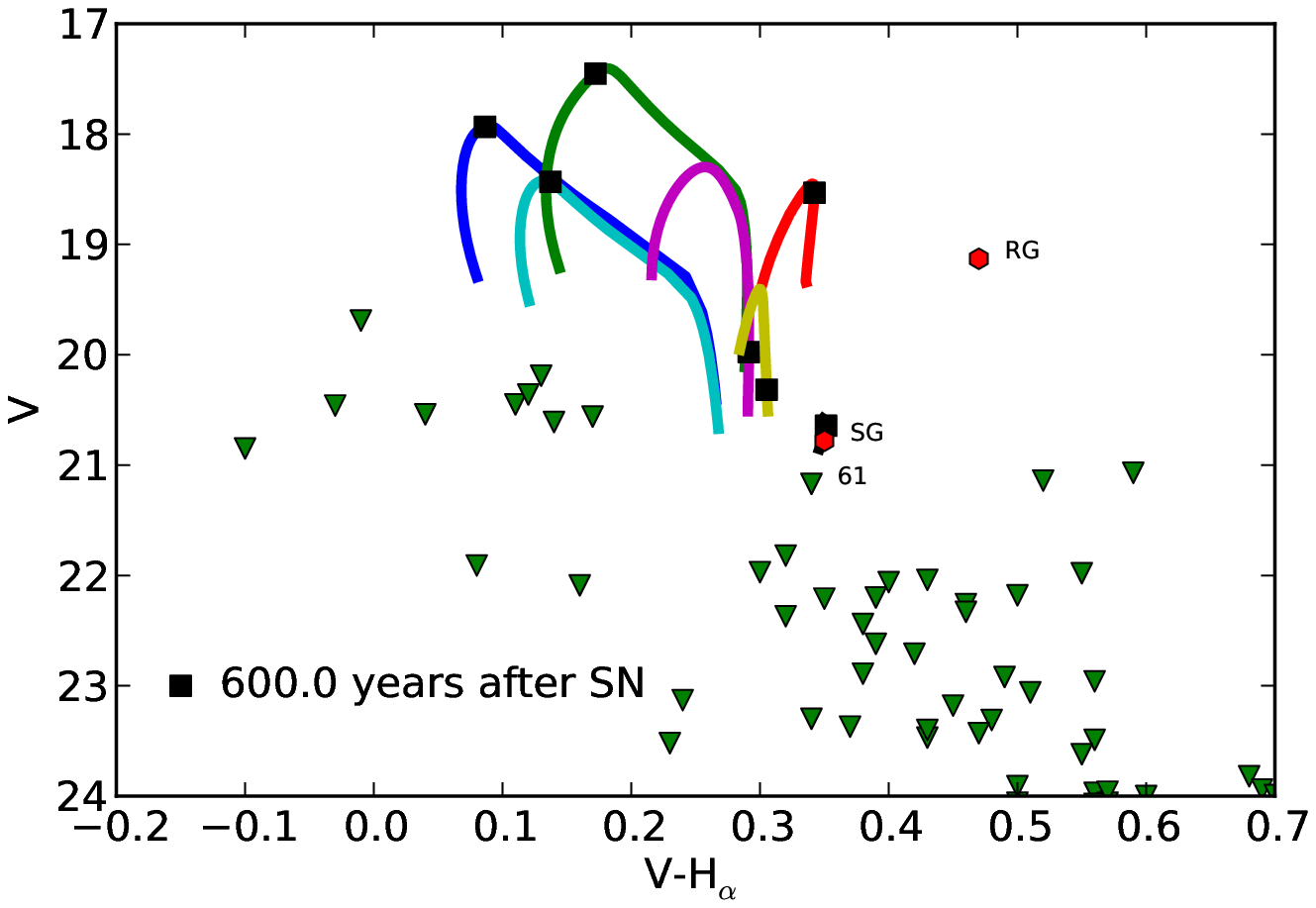}
\end{center}
\caption{\label{fig_snrs}
\scriptsize
Evolutionary tracks in effective temperature vs.\ surface gravity diagrams
and H-R diagram of SC candidates in SN~1572 (top), SN~1006 (middle), and SNR~0519-69.0 (bottom).
Color lines indicate the simulated evolutionary tracks for different SC models.
Black squares represent the conditions at the ages of Ia~SNRs.
For SNR~0519-69.0, an extinction of $A_V=0.42$ is adopted in our SC calculations
\citep{2004AJ....128.1606Z}.
Note that the ages of historical recorded supernovae (SN~1572, SN~1006) are very accurate,
but the ages of SNRs in LMC are less accurate.
Fortunately, the early evolution of MS-SCs and late evolution of He-SCs after a hundred years are relatively slow.
Other symbols denote the observational conditions of SC candidates from different papers
(SN~1572: green up-triangles, \citealt{2009ApJ...691....1G} and red circles, \citealt{2013ApJ...774...99K};
SN~1006: red pentagons, \citealt{2012Natur.489..533G} and green stars, \citealt{2012ApJ...759....7K}; and
SNR~0519-69.0: green down-triangles and red circles, \citealt{2012ApJ...747L..19E}).}

\end{figure}

\cite{2004Natur.431.1069R} found a subgiant star, namely Tycho~G,
characterized by a high radial velocity $\sim 108 \pm 6$~km~s$^{-1}$ that could be
related to the original orbital speed of the white dwarf companion in the SDS.
\citet[][hereafter K13]{2013ApJ...774...99K} recently updated this radial velocity to $v_{\rm LSR} \sim 79$~km~s$^{-1}$,
which is still anomalous for the region.
Figure~\ref{fig_snrs} shows our predictions of post-impact conditions of MS-SCs compared with
observations by \citet[][hereafter GH09]{2009ApJ...691....1G} and K13.
GH09 reported that Tycho G is a subgiant with effective temperature
$\Teff = 5900 \pm 100$~K, surface gravity $\log g=3.85 \pm 0.3$~dex, and [Fe/He]~$=-0.05 \pm 0.09$~dex,
using Keck high resolution optical spectra.
The distance of Tycho~G is also comparable to that of Tycho's SNR.
Using the same data but with a slightly different analysis tool,
K13 determined hotter and less luminous characteristics for Tycho~G
($\Teff = 6000 \pm 100$~K, $\log g=4 \pm 0.3$~dex, and [Fe/H]$=-0.13 \pm 0.13$~dex;
see K13 for a more detailed comparison of their results with GH09).

Our Model E ($\Teff = 5737$~K, $\log g=3.3$~dex,
and $L=20.2 L_\odot$; see Table~3 in P12b) has a similar effective temperature
to that of Tycho G, but is brighter than the luminosity reported by GH09 
($L \sim 1.9-7.6 L_\odot$, assuming a mass of $1M_\odot$).
We note that our Model E is more massive ($M_E=1.44 M_\odot$)
than the mass ($1M_\odot$) assumed in GH09 and K13, and the stellar radius
and evolution stage are also different with what is suggested in \citet[][hereafter B14]{2014MNRAS.439..354B}.
The evolution of SCs after the SN explosion depends not only on the amount of mass lost and energy deposition,
but also on the depth of energy deposition. Therefore, a less massive companion, different evolutionary stage,
asymmetric explosion, or sub-Chandrasekhar mass explosion may better match Tycho~G.
If Tycho~G is indeed similar to our Model~E, we predict a magnitude change of $\sim 0.01$~mag for 
every 10 years (Figure~\ref{fig_mag}).

B14 determined high-accuracy proper motions for $872$ SC candidates,
including Tycho~G, in the central region of Tycho's SNR.
They obtained a tangential velocity for Tycho~G of $v_t \sim 64\pm 11$~km~s$^{-1}$.
Together with the observed radial velocity (80~km~s$^{-1}$) of Tycho~G and the average radial velocity
(37~km~s$^{-1}$) in the direction of Tycho's SNR from the Sun,
the linear velocity with respect to the center of Tycho's SNR
can be determined to be $\sim 77 \pm 16$~km~s$^{-1}$, if Tycho~G were the SC (B14).
This value is half that of the linear velocities in our models in Table~\ref{tab_models}.
An inclination angle $i \sim 34^\circ$, orbital separation $a \sim 26 \pm 12 R_\odot$,
and orbital period $P=10\pm 7$~days are also suggested by B14.
These progenitor system data are consistent with the spectroscopic observation by GH09 indicating that
Tycho~G is a G-type subgiant.
However, we note that the linear velocity does not equal the orbital speed in the binary system,
since the SN kick may contribute up to one-third of its final linear speed
(P12a; \citealt{2012A&A...548A...2L}).

\citet[][hereafter K09]{2009ApJ...701.1665K} found an upper limit for the rotational speed of Tycho G of
$v_{\rm rot} \sin i \lesssim7.5$~km~s$^{-1}$, updated to $v_{\rm rot} \sin i \lesssim 6$~km~s$^{-1}$ in K13,
causing them to question Tycho~G as a SC candidate.
P12a, P12b, and \cite{2013A&A...554A.109L} studied the SN~impact and post-impact
conditions of MS-like binary companions and found that this discrepancy can be resolved due to the
loss of angular momentum during the SN impact and post-impact expansion.
However, even when applying the inclination angle in B14,
our best model (model~E) still has a $v_r \sin i =15$~km~s$^{-1}$,
which lies above the upper limit observed by K09 and K13.
Similar results have been found in \cite{2013A&A...554A.109L} as well.
Therefore, a less massive or less compact companion model is required to better fit with observations.
Fortunately, using the inclination angle and orbital period derived in B14,
a low rotational speed of $v_{\rm rot} \sin i \sim 5.6$~km~s$^{-1}$ can be calculated by assuming tidal locking,
explaining the non-detection of rotational speed in K09 and K13.

On the other hand, \cite{2007PASJ...59..811I} comments
that the absence of an Fe~I line feature at 372~nm in Tycho~G argues against the SC
interpretation as there is no evidence for absorption due to the SN ejecta in the stellar spectrum.
In contrast, Tycho~E shows a strong blueshifted Fe~I absorption line without redshifted lines,
which implies that it is within the remnant
and that its projected position is close to the center of Tycho's SNR,
possibly qualifying it as another SC candidate.
However, K13 suggests that Tycho E is far behind the SNR
and hence the low column density on the receding side of the remnant could explain
the lack of redshifted lines.
The recent observation of Tycho~E reveals an effective temperature $\Teff=5825$~K and
surface gravity $\log g = 3.4$~dex (K13), which is also close to our model E ($\Teff = 5737$~K and $\log g=3.3$~dex).
However, the low radial velocity ($v_{\rm LSR} \sim 55.91\pm0.27$~km~s$^{-1}$) and large distance
($d\sim 11.2\pm7.5$~kpc, in comparison to $d_{\rm Tycho} \sim 2.8 \pm 0.8$~kpc)
make it less likely to be the SC, although the distance uncertainty is large.
Finally, GH09 also have suggested that Tycho~E could be a double
lined binary, which does not provide support for it as a SC.
Further detailed observations of these SC candidates are necessary
to establish whether any of them is connected to Tycho's SN.

\subsection{SN~1006}

The lack of a compact remnant star, and the amount of iron observed inside the SNR
\citep{1997ApJ...481..838H} 
indicate that SN~1006 was a SN~Ia.
Measurements of its proper motion indicate that
SN~1006 is the closest historical Ia~SNR with a distance $d\sim 2.18 \pm 0.08$~kpc \citep{2003ApJ...585..324W} and an age of $t_{\rm SNR} = 1008$~yr.
The geometric center of the remnant is also well determined in X-ray and radio \citep{2003ApJ...585..324W}.
Furthermore, the foreground extinction is low due to its high Galactic latitude ($b=14.6$),
providing a good environment in which to search for SC candidates.

Searches for SC candidates in SN~1006 have been done
using two independent observations by \citet[][hereafter GH12]{2012Natur.489..533G}
and \citet[][hereafter K12]{2012ApJ...759....7K}.
Both teams suggest that there is no evidence for the survival of a SC in the
central region of SN~1006.
K12 observed all stars to a limit of 0.5~$L_\odot(V)$ at the distance of SN~1006
within the central 2 arcminutes. They found no stars as bright as predicted in
\cite{2000ApJS..128..615M} and \cite{2003astro.ph..3660P},
and no stars show significant rotation.
Similar conclusions have been drawn by GH12 using observations of the central 4 arcminutes.

However, we note that the stars B90474 ($\log g=3.05 \pm 0.12$~cm~s$^{-2}$,
$\Teff= 5051 \pm 38$~K) and B14707 ($\log g =3.36 \pm 0.15$~cm~s$^{-2}$, $\Teff=5065\pm 47$~K) in
GH12 have surface gravities similar to our Model G ($\log g = 3.14$~dex, $\Teff=
5288$~K), although with lower effective temperatures (see Figure~\ref{fig_snrs}).
Furthermore, the star B90474 has a high radial velocity ($v_{\rm rad} = 98.10 \pm 1.95$~km~s$^{-1}$).
However, unlike the case of Tycho's SN, SN~1006 lies 500~pc above the Galactic plane, and
therefore, the radial velocities of surrounding stars do not exhibit a simple trend.
In addition, B90474 has a larger distance uncertainty ($d \sim 4.78 \pm 2.0$~kpc)
and B14707 ($d\sim 1.37 \pm 0.58$~kpc) lies at a much closer distance than the SNR.

In both GH12 and K12, the authors use the geometric center of the X-ray and radio observations.
The error circles included in GH12 and K12 are $4'$ and $2'$ respectively,
corresponding to a SC moving with a speed of $\sim 1,000$~km~s$^{-1}$ for 2,000 and 1,000 yr
at a distance of 2.2~kpc.
\cite{2005ApJ...624..189W} have suggested that there is an offset between the geometric
center and the center of the iron core.
Furthermore, \cite{2013ApJ...771...56U} and \cite{2014ApJ...781...65W} have shown that
the ejecta distribution in SN~1006 is asymmetric and concentrated in the SE quadrant,
suggesting a $5' \sim 3.2$~pc offset to the geometric center \citep{2013ApJ...771...56U}.
\cite{2014ApJ...781...65W} reported that the expansion velocity varies significantly with
azimuth ($\sim 3000$~km~s$^{-1}$ in the NW and $7400$~km~s$^{-1}$ in the SE).
Therefore, the error circles used in GH12 and K12 may not include
the real explosion center.

On the other hand, the Schweizer-Middleditch star is a subdwarf OB (sdOB) star which is located at the
center of SN~1006 and has strong Fe absorption lines \citep{1980ApJ...241.1039S}.
It is relatively bright ($V=16.74$, $B-V=-0.14$),
with low foreground extinction ($E(B-V)=0.1$; \citealt{1980ApJ...241.1039S}).
It is consistent with the He-SC models; however, the presence of redshifted absorption lines due to
supernova ejecta suggests that it is more likely a background star \citep{1983ApJ...269L...5W}.

\subsection{SN~1604 (Kepler's SNR)}

Kepler's SNR has been identified as a Ia~SNR based on X-ray observations of its O/Fe ratio
\citep{2007ApJ...668L.135R},
but its distance is still uncertain ($d\sim 4-6.4$~kpc; \citealt{2012A&A...537A.139C}).
The interaction of the supernova ejecta with circumstellar material in Kepler's SNR provides
evidence that Kepler's SNR may have originated from the SDS in an evolutionary channel
consisting of a WD and an asymptotic giant branch (AGB) star
\citep{2012A&A...537A.139C, 2013ApJ...764...63B}.
However, the circumstellar medium could be also explained by  
the stellar wind from a massive progenitor in core-collpase supernova.
The low iron mass of $0.01 M_\odot$ \citep{1990PASJ...42..279H} and  
a progenitor mass of $7M_\odot$ \citep{1985ApJ...291..544H} 
suggest that Kepler's SN may not be a SN~Ia.

Recently, \citet[][hereafter K14]{2014ApJ...782...27K} have ruled out red giants as SCs due to the lack
of bright stars in the central SNR.
The observed radial velocity of SC candidates also shows only a small possibility of being associated with
the SN explosion.
However, 24 stars with $L_V > 10 L_\odot$
at the center of Kepler's SNR have been found by K14, and five of them have $L_V > 20 L_\odot$,
perhaps suggesting a relatively higher probability of being a SC.
Additional observations are required to test these candidates.

As we pointed out in \S~2, nearly all the envelope of a RG should be removed during the SN impact
in the RG donor channel, and this is likely to be the case for the AGB donor channel as well.
Therefore, the SC in Kepler's SNR would cease to be a AGB star, 
which would explain the non-detection of a giant star in K14.
Follow-up observations and a detailed study of the evolution of the AGB(RG) channel
in the SDS and its SCs will be necessary in order to understand the progenitor system of Kepler's SNR.

\subsection{RCW~86}

RCW~86 is the oldest known Galactic SN~Ia.
It was recently identified as a Ia~SNR by \cite{2011ApJ...741...96W}
and \cite{2011PASJ...63S.837Y}, who estimate that the integrated
Fe-K emission corresponds to a total Fe mass of about $1M_\odot$.
Its possible association with SN~185 is somewhat dubious since it 
would imply the age of RCW~86 is 1,829~yr.  The large size of its radius at an estimated distance
of $\sim 2.5$~kpc \citep{2011ApJ...741...96W}
(or $d \sim 2.8$~kpc; \citealt{1996A&A...315..243R})
suggests a very high shock speed ($\sim 7,800$~km~s$^{-1}$) or a much older age.
One explanation is that it originated in a cavity explosion \citep{1997A&A...328..628V, 2007ApJ...662..472B},
although cavity explosions are more common in core-collapse SNe. 
Hence, the connection between RCW~86 and SN~185 has yet to be established.
In addition, the derived ambient density ($0.075$~cm$^{-3}$) found by \cite{2011PASJ...63S.837Y}
suggests that an unusually low-density
cavity surrounds the SNR.  This can be understood either as resulting from the existence of a
strong stellar wind from the progenitor itself or as an outflow from the nearby OB association
discovered by \cite{1969AJ.....74..879W}.  If RCW~86 was a member of this group, the age of this
group may place some constraints on the delay-time of this SN~Ia, although the ages of these
OB stars are still unknown. The wind-blown bubble scenario by \cite{2011ApJ...741...96W} suggests
that RCW~86 originated in the SDS.

Therefore, a search for SC candidates in RCW~86 seems reasonable.
However, the two expansion fronts in the southwest and the northeast
make it difficult to measure the center of the explosion (K14).
In our calculations, all MS-SC models reach the brightest phase ($> 100L_\odot$)
and highest effective temperature at around $500-3000$~yr, depending on the model.
Therefore the nearby distance and low extinction ($A_V \sim 1.7$; \citealt{1983MNRAS.204..273L})
make it easier to distinguish SC from unrelated stars.

\section{Type Ia SNRs in the Magellanic Clouds \label{sec_lmc}}

The Magellanic Clouds (MCs) are excellent environments to search for SCs
for several reasons. Ia~SNRs in the Large Magellanic Cloud (LMC) and
Small Magellanic Cloud (SMC) are at known distances, 50~kpc and 60~kpc,
that are sufficiently close to detect SCs with HST based upon the results in \S~2.
The MCs are nearly face-on galaxies, minimizing confusion along the line of sight
\citep{2001AJ....122.1807V, 2012ApJ...744..128S}.
The foreground Galactic extinction is low, and the internal extinction of the MCs is modest
\citep{2011AJ....141..158H}.
Finally, the star formation history (SFH) and metallicity of the MCs are different than
those of the Milky Way \citep{2004AJ....127.1531H, 2009AJ....138.1243H, 2010MNRAS.407.1314M},
allowing the possibility of testing SN~Ia progenitors in different type of galaxies \citep{2013Sci...340..170W}.

Ten Ia~SNRs have been reported to lie in the LMC and four in the SMC (see Table~\ref{tab1}).
Since only four Galactic Ia~SNRs have been identified,
these fourteen Ia~SNRs provide a larger sample of Ia~SNRs to search for SCs.
In addition, the distances of Ia~SNRs in the LMC and SMC are known more accurately than for Galactic
Ia~SNRs, which reduces the uncertainties in comparing observations with the theoretical
predictions.

Recently, the youngest Ia~SNR in the LMC, SNR~0509-67.5,
has been studied by \cite{2012Natur.481..164S}.
SNR~0509-67.5 formed from a SN~1991T-like SN Ia $\sim 400$~yrs ago.
The observation of the central region of SNR~0509-67.5 within a radius
of $1.43''$, corresponding to a distance of 0.36~pc from the center of SNR~0509-67.5,
shows no star brighter than $V=26.9$ ($M_V=8.4$) in this region.
This result rules out all standard SDS channels and suggests that
SNR~0509-67.5 originated in the DDS \citep{2012Natur.481..164S}.

\cite{2012ApJ...747L..19E} studied the central sources within an error circle of $4.7''$ radius about the center the
$\sim 600$~yr old Ia~SNR 0519-69.0 using HST.
They found 27 MS stars brighter than $m_V =22.7$ magnitude.
This result requires the progenitor of SNR~0519-69.0 to arise from either the DDS
or the SDS with a supersoft source.
We point out that \cite{2012ApJ...747L..19E} found a sub-giant star in SNR 0519-69.0
($V = 20.78 \pm 0.01$~mag, and $V-H_\alpha=0.35 \pm 0.01$).
Although this star is consistent with our model G star
($V = 20.7$~mag, and $V-H_\alpha=0.35$; the black line in Figure~\ref{fig_snrs}),
it lies close to the edge of the possible error circle, which suggests that
it may not be the SC of SNR~0519-69.0 unless the explosion was asymmetric.

\section{Other Nearby Galaxies \label{sec_galaxies}}

Besides the LMC and SMC, other nearby galaxies can provide samples that probe diverse
environments for SCs.  In particular, the metallicities and SFHs in other nearby galaxies
differ from those of the Milky Way and Magellanic Clouds, which could affect pre-SN conditions,
the SN~Ia explosion itself, and, potentially, the post-impact evolution.

In Section~\ref{sec_predictions}, we have shown that our predicted MS-SCs (He-SCs) span a range of
absolute magnitude $3 < M_V \lesssim -1$ ($4 \lesssim M_B \lesssim -4$) in F555W (F438W).
By using HST/WFC3 (with U, B, V, and I filters),
the limiting magnitude with S/N of 10 for point sources is $\sim 27$
with a one hour exposure, and $\sim 29$ for a 10 hour exposure,
giving maximum distance moduli of $24-28$ and $23-31$ respectively.
This corresponds to a maximum detectable distance of $0.6-4$~Mpc ($0.4-16$~Mpc)
if there is no extinction.
However, as extinction varies from galaxy to galaxy, the maximum detectable distance would be
smaller in extreme cases.

Given that He-SCs reach their maximum brightness at $10-30$~yr after SN explosions and then
fade within $\sim 100$~yr (P13), their $B$ magnitudes are likely less
than 1 for historic Ia~SNRs.  This yields a distance limit of $d_{\rm SN, max} \lesssim 0.4$~Mpc.
However, the short timescale ($10-30$~yrs) of their luminous phase provides the possibility to detect
slow transitions of their brightness at the locations of recent nearby SNe~Ia with $d_{\rm SN, max}
< 16$~Mpc.  We note that the timescale for the decay of the light curve from the supernova
may take longer than the luminous phase of the SC and, hence, that the maximum detectable distance
could be smaller.

SNRs in nearby galaxies,
including the Andromeda Galaxy
(M31 or NGC~224; \citealt{1981ApJ...247..879B,2012A&A...544A.144S}),
NGC~300 \citep{2012SerAJ.184...19M}, M82 \citep{2010MNRAS.408..607F},
NGC~4214 \citep{2010Ap&SS.330..123D}, NGC~5204, NGC~5585, NGC~6946, M81,
and M101\citep{1997ApJS..112...49M}, have been studied recently.
Therefore, Ia~SNRs in these galaxies should be targeted for SC searches.
However, it is unclear whether these SNRs are Ia~SNRs.

\subsection{Andromeda Galaxy (M31)}

More than 26 SNRs (including Type~I and Type II) have been identified in the Andromeda Galaxy at 0.79~Mpc
by \cite{1981ApJ...247..879B} and \cite{2012A&A...544A.144S}.  M31 has a distance modulus of 24.5, which
makes it possible to detect all of our MS-SCs and most He~SCs using HST/WFC3.  The Panchromatic
Hubble Andromeda Treasury (PHAT) multi-cycle program has observed about one third of M31 using HST
\citep{2012ApJS..200...18D}.  Its UVIS data reached a magnitude limit of $\sim 25$ in the F275W and F336W
bands; ACS data reach maximum depths of $\sim 28$ magnitudes in F475W and $\sim 27$ magnitudes in
F814W in the uncrowded outer disk.  In these same regions,
WFC3/IR data reach maximum depths of $\sim 26.5$ and $\sim 25.5$ in F110W and F160W, respectively.

\cite{2012A&A...544A.144S} have listed 26 X-ray SNRs and 20 X-ray SNR candidates in M31 based on their
X-ray, optical, and radio emission, which is the most recent complete list of X-ray SNRs in M31.
Therefore, using their list together with PHAT's data, it may be possible to identify SC candidates.

SN~1885a (S~Andromedae) is a subluminous SN~Ia in M31 \citep{2002AJ....123.2045V, 2007ApJ...658..396F}.
Its young age and the small size of its remnant make it easier to search for a SC.
However, the surrounding light is dominated by the bulge of M31, making SC searches difficult unless the SC
is overluminous.

\subsection{Recent nearby SNe~Ia}

SN~2014j was discovered by Stephen Fossey and his students on January 21, 2014 and reported
as a SN~Ia on January 22, 2014 \citep{2014CBET.3792....1F, 2014ATel.5786....1C}.
SN~2014j is the closest SN~Ia ($d \sim 3.5 \pm 0.3$~Mpc; \citealt{2006Ap.....49....3K})
discovered in the past 42 years.
The next close one is SN~1972e (in NGC~5253)
with a distance of $4.64 \pm 0.7$~Mpc \citep{1992A&A...257L...1D}.
SN~2011fe is also a nearby SN~Ia discovered in the modern astronomical era.
However, the distance of SN~2011fe (in M101; $d \sim 6.4$~Mpc, \citealt{2011ApJ...733..124S})
is above our predicted maximum distance for MS-SCs.
A search for SCs in SN~2011fe will be very challenging.
Furthermore, the emptiness of the surrounding medium \citep{2012ApJ...750..164C}
and the lack of pre-explosion optical and X-ray source \citep{2012ApJ...749..141L}
suggest a DDS progenitor for SN~2011fe. 

SN~2014j is almost twice as close as SN~2011fe, and
the distance (3.5~Mpc) is within our predicted detection limit (4~Mpc) for MS-SCs,
though the distance is close to the limit.
HST archival observations do not rule out SDS progenitors such as
recurrent novae or some classical novae \citep{2014ATel.5824....1C, 2014ATel.5849}.
However, even when the SNR becomes transparent, any MS-SC would need a few hundred years
to become bright enough to be detected by HST.
If the SC is a He-SC, a bright luminous OB-like star is expected to be observed in the next few decades.

\section{Conclusions}

We have determined the colors and magnitudes of seven MS- and four He-SC models based on our stellar evolutionary
calculations (P12b, P13) and predicted their rates of change as functions of time.
We have also studied the linear and rotational speeds of SCs and the potential ejecta contamination of SCs.
Comparisons of the model predictions for SCs with the Galactic SNe 1572 and 1006 are presented,
assuming all the candidate stars have the same distance as their host SNRs.
In particular, it is found that both Tycho~E ($\Teff=5825$~K, $\log g = 3.4$~dex)
and Tycho~G ($\Teff = 5900 \pm 100$~K, $\log g=3.85 \pm 0.3$~dex)
approximately fit our predicted MS-SC Model~E
($\Teff = 5737$~K, $\log g=3.3$~dex),
although there are small discrepancies (K13).  Furthermore, two sub-giants
in the error circle of SN~1006 are also found to have magnitudes consistent with our MS~SC
Model~G ($\Teff= 5288$~K, $\log g = 3.14$~dex),
but have lower effective temperatures \citep{2012Natur.489..533G}.
However it should be noted that the uncertainties in the distances of these SC candidates are large,
and that the derived effective temperatures and surface gravities are not consistent in different papers.
In addition to the Galactic Ia~SNRs, a sub-giant in the SNR~0519-69.0 is consistent with the magnitude and
color of our Model~G, but the projected position in the SNR is too far from the center
\citep{2012ApJ...747L..19E}.
If such a candidate is a confirmed SC, then the original orbital speed (plus SN kick) would be much
higher than expected unless the explosion was asymmetric.  Although higher orbital speeds would be
expected for He~SCs, there is no evidence for such a candidate.

Based on the current sample of SNe~Ia companion searches,
it is more likely that most have originated from the DDS or peculiar SDS channels.
Peculiar SDS channels such as the M-dwarf channel \citep{2012ApJ...758..123W}
or the spin up/down channel \citep{2011ApJ...730L..34J, 2011ApJ...738L...1D,2012ApJ...759...56D}
may explain the non-detection of SC candidates in Ia~SNRs.
To obtain a better understanding of the progenitor systems of SN~Ia,
we encourage further SC searches in other Galactic or nearby
extragalactic Ia~SNRs.  Unlike Galactic Ia~SNRs, which
have large distance uncertainties, the distances of extragalactic Ia~SNRs are relatively well-known.
We predict that the maximum detectable distance of MS-SCs (He-SCs) is $0.6-4$~Mpc ($0.4-16$~Mpc),
if the apparent magnitude limit is 27 with no extinction, suggesting that the LMC, SMC, and M31
are excellent targets in which to search for SCs.
Furthermore, we also predict He-SCs will not only show high luminosity and effective temperature 
(luminous OB-like stars), but also show a rate of magnitude in B of $\sim 0.1 -1$~mag~yr$^{-1}$ 
and a rate of change of effective temperature of $-1,000-4,500$~K~yr$^{-1}$ in the first decade since explosion.
Future observations of He-SC candidates should look into their magnitude and color changes as well.

Similar analysis can be applied to core-collapse (CC) SNe as well.
The observed high fraction of SN~Ib/c cannot be explained by single stellar evolution,
suggesting that a high fraction of CCSNe were in binary systems \citep{2011MNRAS.412.1522S, 2009ApJ...707.1578K}.
However, the ejected-mass-and-energy and companion types and separations are very different
from those in SN~Ia. In addition, the companion does not need to be in Roche lobe overflow
at the time of explosion. Therefore, the evolution of SCs in CCSNe could be very different
and vary from case to case. The impact of CCSNe on binary companions and their subsequent evolution
are important future work in this area.


\acknowledgments

We thank the anonymous referee for his/her valuable comments and suggestions.
KCP thanks You-Hua Chu for useful discussions about Ia~SNRs in the LMC.
This work was supported by the Computational Science and Engineering (CSE) fellowship
at the University of Illinois at Urbana-Champaign, by the Theoretical Institute for Advanced Research in
Astrophysics (TIARA) in the Academia Sinica Institute of Astronomy and Astrophysics (ASIAA),
and by the European Research Council (ERC) grant FISH.
The FLASH simulations presented here were carried out using the NSF XSEDE Ranger system at the 
Texas Advanced Computing Center under allocation TG-AST040034N.  
FLASH was developed largely by the DOE-supported ASC/Alliances Center for Astrophysical 
Thermonuclear Flashes at the University of Chicago.  



\clearpage

\begin{deluxetable}{lccccc}
\tabletypesize{\scriptsize}
\tablecaption{The MS and He progenitor models \label{tab_models}}
\tablewidth{0pt}
\tablehead{ \colhead{Model}
& \colhead{$M_{\rm b}^\dagger$ ($M_{\odot}$)} & \colhead{$R_{\rm b}$  ($R_{\odot}$)}
& \colhead{$M_{\rm a}^\ddagger$ ($M_{\odot}$)} & \colhead{$R_{\rm a}$  ($R_{\odot}$)}
& \colhead{$v_{\rm linear}$ (km s$^{-1}$)}
}
\startdata
A (MS) & 1.88 & 1.25 & 1.64 & 3.87 & 179 \\
B (MS) & 1.82 & 1.50 & 1.65 & 4.76 & 179 \\
C (MS) & 1.82 & 2.63 & 1.56 & 7.61 & 136 \\
D (MS) & 1.63 & 1.19 & 1.43 & 3.42 & 188 \\
E (MS) & 1.59 & 1.42 & 1.44 & 3.91 & 191 \\
F (MS) & 1.55 & 1.97 & 1.30 & 4.09 & 143 \\
G (MS) & 1.17 & 0.79 & 0.93 & 4.45 & 271 \\
HeWDa (He) & 0.697 & 0.091 & 0.656 & 0.390 & 734 \\
HeWDb (He) & 0.803 & 0.158 & 0.748 & 0.565 & 550 \\
HeWDc (He) & 1.007 & 0.194 & 0.962 & 0.809 & 509 \\
HeWDd (He) & 1.206 & 0.231 & 1.126 & 1.041 & 446 \\
\enddata
\tablecomments{Seven MS progenitor models (A-G) and four He progenitor models (HeWDa-d) are
considered in this study.}
\tablenotetext{\dag}{The mass ($M_{\rm b}$) and radius ($R_{\rm b}$) for different progenitors at the time of the SN explosion.}
\tablenotetext{\ddag}{The initial relaxed hydrostatic mass ($M_{\rm a}$) and radius ($R_{\rm a}$) for different progenitor stars after SN explosion. }
\end{deluxetable}

\clearpage

\begin{deluxetable}{lcrr}
\tabletypesize{\scriptsize}
\tablecaption{Type Ia SNRs in the Milky Way and Magellanic Clouds  \label{tab1}}
\tablewidth{0pt}
\tablehead{
\colhead{Name} & \colhead{Age (yrs)} & \colhead{Distance (kpc)} & \colhead{Radius (pc)}
}
\startdata
RCW~86 (SN~185?)$^\dagger$ & $> 1,800$ & $\sim 2.3-2.8$ (MW)  & $\sim 15$ \\
SN~1006 & 1,007 & $2.2 \pm 0.08$ (MW) &9.3 \\
SN~1572 (Tycho's) & 441 & $2.8 \pm 0.8$ (MW) & 3.8 \\
SN~1604 (Kepler's) & 409 & $\sim 6$ (MW)& 3 \\
B0509-67.5  & $400 \pm 50$  & 50 (LMC)& 3.6 \\
N103B & $1,000-2,000$ & 50 (LMC)& 3.6  \\
B0519-69.0 & $600 \pm 200$ & 50 (LMC) & 3.9  \\
DEM~L71    & $\sim 4,500$  & 50 (LMC)& 8.6 \\
B0548-70.4  & $\sim 7,000$ & 50 (LMC)& 12.5 \\
DEM~L316A & ? & 50 (LMC)& 15 \\
B0534-69.9  & $\sim 10,000$ & 50 (LMC) & 16 \\
DEM~L238 & $10,000-15,000$ & 50 (LMC) & 21  \\
DEM~L249 & $10,000-15,000$ & 50 (LMC) & 23  \\
B0454-67.2 & $\sim 30,000$ & 50 (LMC) & 27  \\
IKT~4  & ? & 60 (SMC)& $\sim 12$ \\
IKT~5  & ? & 60  (SMC)& 15  \\
IKT~25  & ? & 60 (SMC)& 18  \\
DEM~S128 & ? & 60 (SMC) & 26  \\
\enddata
\tablecomments{Four Galactic (MW) Ia~SNRs (SN~1006, SN~1572, SN~1604, and RCW~86)
and fourteen Ia~SNRs in the LMC and SMC that are listed in \cite{2012A&ARv..20...49V}.}
\tablenotetext{\dag}{References: \cite{1969AJ.....74..879W,1996A&A...315..243R}, and \cite{2011ApJ...741...96W}.}
\end{deluxetable}

\clearpage


\end{document}